# DEFINING PLANETS BY A GENERATING FUNCTION

Ivan Kotliarov

St. Petersburg State University

Faculty of Geography

Abstracts: a law of semi-major axes in the Solar system is described. The proposed law is recommended as an additional criterion for planet definition as it represents a generating function of planetary distances.

INTRODUCTION

There are many problems waiting for a final solution in the astronomy of the Solar system, but two of them are particularly interesting: the law of distribution of planetary distances (or, to be more correct, of semi-major axes) and a precise definition of a planet. The seeming simplicity of formulation of these problems and their importance for the science attracts both professional and non-professional astronomers (the situation that clearly reminds us about the famous Fermat's last theorem). Here is the short overview of results obtained for these problems:

1. The most famous formula of planetary distances is undoubtedly the Titius-Bode law (TBL):

$$a_i = 0.4 + 0.3 \times 2^n, \qquad (1)$$

$a_i$ – semi-major axis of the $i$-th planet (counting from the Sun), AU;

$n$ – exponent, $n = -\infty$ for Mercury and $n = i - 2$ for all other planets.

TBL is purely phenomenological – it has no explanation based on celestial mechanics. Moreover, TBL is not valid for Neptune; it *may* be considered as correct for Pluto, but if and only if we postulate for Pluto $n = i - 3$.

Another objection is the unclear nature of the exponent *n* which is equal to $-\infty$ for Mercury. Obviously it is difficult to ascribe any physical value for this exponent in the case of Mercury.

Astronomers tried to find a dynamical explanation for TBL but eventually failed (Graner, Dubrulle 1994). Their attempts to fit all the planets (including Pluto) into one formula (generally into an exponential one) were unsuccessful, too – either the discrepancies between the observed and the calculated distances were too high or these formulae supposed the existence of not yet discovered planets. A detailed overview of older attempts can be found in (Nieto 1972), (Badolati 1980) and (Badolati 1982). New publications on this subject include, for example, an essay to fit all "major" bodies of the Solar system (i. e. planets and dwarf planets according to the 2006 IAU resolution) into a Titius-Bode-like formula is outlined in (Ortiz et al. 2007). Lubos Neslusan (2004) performed a statistical analysis of the Titius-Bode distribution and showed that it is not due to a chance. In a recent publication Arcadio Poveda and Patricia Lara (2008) extrapolated TBL (in the exponential form) to extra-solar planetary system and even predicted on a basis of their hypothesis At last, it is necessary to list the important publication (Santiago Alcobé and Rafael Cubarsi 2006) where the authors applied a Titius-Bode like formula to the distribution of stellar populations;

2. According to the 2003 IAU resolution a planet of the Solar system is defined as a celestial body orbiting the Sun, massive enough to be rounded by its own gravity and having cleared its neighborhood. If a celestial body fulfills only first two criteria it is considered to be a dwarf planet. The list of planets includes Mercury, Venus, the Earth, Mars, Jupiter, Saturn, Uranus, Neptune. Ceres, Pluto and Eris were classified as dwarf planets. This definition is obviously controversial, and astronomers tried to fix it. The most interesting attempt in my opinion was made by Steven Soter (2006) – most notably, the author introduced a quantitative measure for "clearing the neighborhood" and proposed a transparent definition of the orbital zone. However, his solution is not flawless either as he virtually does not take into account the dynamical characteristics of the planets and the evolution of these characteristics.

Interestingly enough, to the best of my knowledge nobody has tried yet to combine these two problems into one. Below I will try to demonstrate that this combination may give us a strict solution of both these problems on a purely mathematical ground.

AN AMENDMENT OF THE TITIUS-BODE LAW

It is logical to use TBL as a basis for a new formula of distribution of planetary distances. But first of all we have to choose the form of TBL we are going to use. There are two main forms for this law – the classical one (formula (1) above) and the exponential one (the latter has several variants). As it was explained earlier, the exponential formulae do not provide us with a good fit between observed and calculated distances and high discrepancies exist for all planets. Therefore, in my opinion, exponential formulae should be discarded.

The main problems with the classical formula for TBL are the discrepancies for Neptune, Pluto and Eris and the fact that the exponent $n$ is physically meaningless. Let us try to find a basis for the exponent.

Actually, it would be more correct to write down the traditional formula for TBL (1) as follows:

$$a_i = 0.4 + 0.3 \times 2^N, \qquad (2)$$

$$N = F(i), \qquad (3)$$

$i$ – orbital number (the number of the planet counting from the Sun).

That is, $N$ is a function of the orbital number, but not the orbital number itself.

As $N = -\infty$ for Mercury and $N = n - 2$ for all other planets I would propose the following explicit presentation of $N = F(n)$:

$$N = \frac{n-2}{\text{sign}(n-1)}. \qquad (4)$$

In this case the orbital number for Mercury is 1 – as it is expected to be which is perfect from the logical point of view, but $N = -\infty$ as it is prescribed by TBL in its traditional form.

It is interesting to mention that the formula (4) – while being absolutely obvious – has not been proposed by anybody. Of course, it does not provide the exponent $N$ with a physical meaning as the orbital number is not a physical characteristic of a planet according to the mainstream astronomy. However, it links the unexplained exponent $N$ in the formula (2) with the orbital number of the planet which makes the TBL in its traditional form less esoterical.

BUTUSOV'S LAW

In 1973 Kirill Butusov, an astronomer from the USSR, proposed a new structural law (Butusov 1973). This law is almost unknown outside Russia as all publications dedicated to it are in Russian (some information of this law can be found in (Kotliarov 2008).

According to this law (I will call it Butusov's law), the product of semi-major axes of planets that are symmetrical related to Jupiter (i.e., Ceres and Saturn, Mars and Uranus etc) is equal to the square of the semi-major axis of Jupiter. It means that the distribution of the planetary distances in the Solar system may be described by the following formula:

$$a_6^2 = a_{6+i} \times a_{6-i}, \qquad (5)$$

$a_6$ – Jupiter's semi-major axis, AU,

$i \in \{0, 1, 2, 3, 4, 5\}$,

$6 \pm i$ – number of a planet counting from the Sun,

$a_{6 \pm i}$ – semi-major axis of the $(6 \pm i)$-th planet (counting from the Sun), AU.

For example, if $i = 2$, then

$$a_6^2 = a_4 \times a_8 \Rightarrow a_{Jupiter}^2 = a_{Mars} \times a_{Uranus}. \qquad (6)$$

Butusov's law simply says that there is a Jovian symmetry of planetary distances in the Solar system (similar results were independently obtained by Sven-Ingmar Ragnarsson (1995)).

Butusov's results are summed up in the table 1.

TABLE 1

BUTUSOV'S LAW

| Planet | $n$ | $i$ | $a$, AU | $A_{6+i} * a_{6-i}$ | $d$, % |
|---|---|---|---|---|---|
| Mercury | 1 | 5 | 0.387 | 26.188 | - 3.264 |
| Venus | 2 | 4 | 0.723 | 28.545 | 5.446 |
| Earth | 3 | 3 | 1.000 | 28.263 | 4.402 |
| Mars | 4 | 2 | 1.524 | 29.247 | 8.038 |
| Ceres | 5 | 1 | 2.766 | 26.379 | - 2.556 |
| Jupiter | 6 | 0 | 5.203 | 27.071 | 0,00 |
| Saturn | 7 | 1 | 9.537 | 26.379 | - 2.556 |
| Uranus | 8 | 2 | 19.191 | 29.247 | 8.038 |
| Neptune | 9 | 3 | 28.263 | 28.263 | 4.402 |
| Pluto | 10 | 4 | 39.482 | 28.545 | 5.446 |
| Eris | 11 | 5 | 67.668 | 26.188 | - 3.264 |

$$d = \frac{a_{6+i} \times a_{6-i} - a_6^2}{a_6^2} \times 100\%.$$

Indeed, Butusov's law is not absolutely precise – but the difference between the distances calculated on its basis is much lower than the difference for calculations based on exponential forms of TBL.

It is important to mention that Butusov's law is also valid for Eris.

Butusov's law – like TBL – is purely phenomenological, that is, it has no recognized theoretical basis. Nevertheless, it is correct for all celestial bodies that are recognized as planets and dwarf planets by the International Astronomical Union (see the table 1 above).

Butusov tried in his later works to provide such a basis, but his results were not accepted by the scholarly community as they were based on non-traditional approaches towards the Solar system. The discussion of this problem is beyond the scope of this paper, we will simply admit that Butusov's law represent a strict phenomenological rule existing in the Solar system. Therefore, we are entitled to use it to build up a formula of distribution of the planetary distances.

## THE COMBINED TITIUS-BODE-BUTUSOV'S LAW AS A GENERATING FUNCTION FOR PLANETS IN THE SOLAR SYSTEM

From the mathematical point of view all objects rotating around the Sun form a set. This set can be divided into several subsets – planets, asteroids, comets, moons, debris etc. According to the IAU resolution, objects are considered to belong to the planetary subset (or the planetary sequence) if they meet the following requirements: they orbit the Sun, are massive enough to be rounded by its own gravity and cleared their neighborhood. However, in the set theory for a given sequence there is a notion of generating function G($n$), $n \in N$, $N$ is the set of natural numbers with the following characteristic:

$$G(n) = b_n,$$

$b_n$ – $n$-th member of the sequence.

The question is if it is possible to design a generating function for planets, or, to be more precise for one of their characteristics that form a sequence? The reply is obviously "yes" as such generating function is obviously the function of planetary distances with respect to the orbital number.

We have two structural laws for the Solar system:

- Titius-Bode law, which proposes a correlation between the planetary distance and the orbital number of the given planet (and is therefore a generating function of planetary distances), but is not correct for Neptune, Pluto and Eris:

- Butusov's law, which is correct for all major bodies of the Solar system (i.e. for planets and dwarf planets, but does not provide us with a correlation between the planetary distance and the orbital number.

The obvious solution is to combine these two laws into one.

Let us introduce an additional parameter $d_n$:

$$d_n = \text{sign}(6 - n), \quad (7)$$

$n$ – orbital number.

This parameter is the characteristic of the Jovian symmetry of the $n$-th planet (it was introduced by Ragnarsson (1995) and independently rediscovered by myself (Kotliarov 2007)).

Let us study the formula (5) for a $n$-th planet, $1 \le n \le 6$. Obviously

$$n = 6 - i, \; i \in \{0, 1, 2, 3, 4, 5\},$$
$$i = 6 - n \Rightarrow 6 + i = 12 - n. \quad (8)$$

Thanks to the formula (8) the formula (5) can be rewritten as follows:

$$a_6^2 = a_n \times a_{12-n}. \quad (9)$$

The formula (9) introduces the actual number of a planet into the Butusov's law. Now we have to fine a precise formula for the planetary distances on a basis of TBL and the modified form of the Butusov's law (9).

It is known that TBL is correct for $1 \le n \le 6$ (actually for $1 \le n \le 8$, but this is not important for us). Therefore

$$a_n = \begin{cases} 0.4 + 0.3 \times 2^{\frac{n-2}{\text{sign}(n-1)}}, & 1 \le n \le 6 \\ \dfrac{5.203^2}{0.4 + 0.3 \times 2^{\frac{10-n}{\text{sign}(11-n)}}}, & 6 < n \end{cases}, \quad (10)$$

where

$$5.203^2 = a_6^2.$$

The formulae (10) are precise (it can be easily checked) and may be considered as a generating function. However, it is logical to simplify them using the parameter $d_n$ introduced above:

$$a_n = 5.203^{1-d_n} \times \left( 0.4 + 0.3 \times 2^{\frac{Z_n - 2}{\text{sign}(Z_n - 1)}} \right)^{d_n}, \quad (11)$$

$$Z_n = n \, \text{Heav}(6 - n) + (12 - n)(1 - \text{Heav}(6 - n)), \quad (12)$$

where Heav($x$) is the Heaviside function:

$$\mathrm{Heav}(x) = \begin{cases} 1, & x \geq 0 \\ 0, & x < 0 \end{cases}.$$

The formula (11) is the strict phenomenological law of distribution of planetary distances and in the same time the generating function for planetary distances. Therefore a planet in the Solar system can be defined as a celestial body that meets the requirements of the IAU resolution and has the semi-major axis defined by the formula (11) in accordance with its orbital number (the sequence of orbital numbers should be a continuous natural row). The definition of a dwarf planet is similar.

POSSIBLE CONTROVERSY

It is easy to see that the formula (11) provides place for just 11 planets in the Solar system (so Eris occupies the last possible planetary orbit) and excludes a potential candidate – Sedna, as it has no counterpart within the orbit of Mercury. This limitation may seem arbitrary.

It is useful to remember, however, that a planet is an important taxonomic category and inflating their number will erase the difference between planets and other celestial bodies. Therefore their number should be limited and the generating function proposed above provides us with an excellent tool for such logical exclusion. Of course, if astronomers manage to find a planetary object within the orbit of Mercury (Vulcan) or to provide us with an uncontestable proof of its existence in the past, additional objects would be qualified to be classified as (dwarf) planets – but on a strictly mathematical basis provided by the formula (11). Otherwise Sedna and all other planetoids outside the orbit of Eris should be excluded from the category "planets" as they do not fit the formula (11). It will help us to maintain the distinction between major bodies of the Solar system (planets and dwarf planets) and other types of objects.

Another objection against the formula (11) may be that it is too complicated. Indeed, it is complicated – but the definitions of meter or second adopted in the metrology are complicated too, but people supports these definitions as they provide us with a clear and unambiguous standards for measurements. The same is true for the formula (11) – it provides us with a clear mathematical definition of a planet.

Obviously, the formula (11) has no dynamical explanation. In addition, it is well known that it is possible to fit a limited number of objects into a formula, so the formula (11) probably has no physical significance. Indeed this is true – but the goal of this formula is to provide a quantitative basis for classification of celestial bodies within the Solar system. Its goal is to satisfy practical needs, not to provide a deep theoretical insight. Moreover, this law (11) has no exceptions and does not contain arbitrary parameters or constants – all variables and constants either have a physical explanation (like the semi-major axis of Jupiter) or are calculated from the orbital number. Therefore, it can be accepted as a law of distribution of planetary distances (and as the generating function of planetary semi-axes) despite the absence of dynamical explanation.

The proposed law can be easily extrapolated to other planetary systems – provided that there exists a similar symmetry and a formula of planetary distances valid for planets with numbers j, $1 \leq j \leq k$ (k is the orbital number of the center of symmetry) can be established. The law in its general form will look as follows:

$$a_j = a_k^{1-d_j} \times F^{d_j}(J(j)), \qquad (12)$$

where

$a_j$ – semi-major axis of the j-th planet, AU,

$F(j)$ – function of planetary distances for planets with the orbital number $j_i$, $1 \leq i \leq k$,

$d_j$ – parameter of symmetry,

$$d_j = \text{sign}(k - j),$$

$$J(j) = j\,\text{Heav}(k - j) + (2n - j)(1 - \text{Heav}(k - j)). \qquad (13)$$

CONCLUSION

The proposed law offers a precise mathematical description of the distribution of semi-major axes of planets in the Solar system. While not being supported by the theory of celestial mechanics, this law is precise from the phenomenological point of view and can be used as a generating function of planetary distances and, therefore, as an additional criterion for planets.